\documentclass[12pt]{article}
\usepackage{amsfonts}
\usepackage{latexsym}
\usepackage{amsmath}
\usepackage{amssymb}

\newcommand{\newsection}{    
\setcounter{equation}{0}\section}
\def\appendix#1{\addtocounter{section}{1}\setcounter{equation}{0}
\renewcommand{\thesection}{\Alph{section}}
\section*{Appendix \thesection\protect\indent \parbox[t]{11.15cm}{#1}}
\addcontentsline{toc}{section}{Appendix \thesection\ \ \ #1}}

\newcommand{\be}{\begin{eqnarray}}
\newcommand{\ee}{\end{eqnarray}}
\newcommand{\bea}{\begin{eqnarray}}
\newcommand{\eea}{\end{eqnarray}}
\newcommand{\ba}{\begin{array}}
\newcommand{\ea}{\end{array}}

\newcommand{\la}{\label}

\def\a{\alpha}
\def\b{\beta}

\def\g{\gamma}

\font\mybb=msbm10 at 11pt

\def\bb#1{\hbox{\mybb#1}}

\def\bH {\bb{H}}

\def\ur{{\underline {r}}}

\def\km{\mathfrak{m}}
\def\kn{\mathfrak{n}}
\def\kp{\mathfrak{p}}

\begin{document}
\begin{titlepage}
\begin{center}
\vspace{5.0cm}

\vspace{3.0cm} {\Large \bf Brane solitons of (1,0) superconformal theories in six dimensions with hypermultiplets}
\\
[.2cm]

{}\vspace{2.0cm}
 {\large
M.~Akyol and  G.~Papadopoulos
 }

{}

\vspace{1.0cm}
Department of Mathematics\\
King's College London\\
Strand\\
London WC2R 2LS, UK\\

\end{center}
{}
\vskip 3.0 cm
\begin{abstract}
We solve the Killing spinor equations of  6-dimensional (1,0) superconformal theories which include hyper-multiplets   in all cases. We show that the solutions
preserve 1,2,3,4 and 8 supersymmetries. We find models with  self-dual string solitons which are smooth and supported by instantons with an arbitrary gauge group, and 3-brane solitons  as expected from the M-brane intersection rules.
\end{abstract}

\vfill
{{\small ~~~~ ~ mehmet.akyol@kcl.ac.uk}

{\small ~~~~ ~ george.papadopoulos@kcl.ac.uk}}

\end{titlepage}

\setcounter{section}{0}
\setcounter{subsection}{0}


\newsection{Introduction}

A consequence of AdS/CFT correspondence is that the field theory dual of $AdS_7\times S^4$ M-theory background is a (2,0) superconformal theory in six dimensions \cite{maldacena}.
So far  an action\footnote{Apart from the well-known problem with self-dual 3-forms, there may not be an action for such a theory but we shall consider superconformal theories with an action
  as a working hypothesis.}  for such a  theory has not been constructed which is local and  6D Lorentz covariant, though there have been suggestions \cite{lp, chu, chu2} which either preserve a subset
of the required symmetries or do not have a general gauge group because of the rigidity in the existence of Euclidean 3-Lie algebras \cite{gp, gg}. Within this context a (1,0) superconformal theory was suggested in \cite{ssw} and later modified in \cite{ssw3} to include hyper-multiplets.
Although  some of these models admit a local action \cite{bandos} and are manifestly classically (1,0)-supercovariant, they  suffer from several pathologies which include the non existence of a ground state and possibly the presence of negative norm states. Nevertheless they exhibit some desirable features like classical superconformal invariance and have smooth  string solitons supported by instantons \cite{ap3}, see also \cite{wolf} for some mathematical aspects. These string solitons  are in accordance with  the M-brane intersection rules \cite{strominger, pktgp}.  So these theories can be thought of  belonging in the same universality  class of theories
as that which is dual to  $AdS_7\times S^4$, and possibly describe  multiple M5-branes.

An exhaustive investigation of the solutions to the Killing spinor equations (KSEs) of the models described in \cite{ssw} has been presented in \cite{ap3}. In particular, the KSEs have been solved in
all cases and the fractions of supersymmetry preserved by the solitons have been identified.  Moreover a class of string solitons solutions have been found in some models
which are smooth and are supported by instanton configurations.

In this paper, we shall extend the investigation of the KSEs to the models of \cite{ssw3} which include hyper-multiplets.
The technique we use to solve the KSEs is based on spinorial geometry \cite{uggp} as it has been adapted in \cite{ap1} to investigate the solutions to the KSEs
of 6-dimensional (1,0) supergravity.  Because of this, we shall not provide  details of the calculation and the proof. Instead, we shall directly state the results for the hyperini KSEs, which are the KSEs
associated with the hyper-multiplets, and refine some of the conditions that appear on the fields. The solution of the gaugini and tensorini KSEs, which are the KSEs associated to the vector and tensor multiplets, is the same as that given in \cite{ap3} and so the analysis will not be repeated here.
Then we shall use the solution of the gaugini and tensorini KSEs in \cite{ap3} and that of the hyperini KSEs presented here to find   new string   and  3-brane solitons in some of the models of \cite{ssw3}. The new string solitons we find are supported by instantons of arbitrary gauge group, they are smooth at a generic point of the instaton moduli space and the string charge is related to the instanton number. The 3-brane solitons
are supported by holomorphic maps of the hyper-scalars and they are complex curves in the hyper-multiplets target space which is a hyper-K\"ahler cone.
The existence of such solitons are in agreement with expectations from the M-brane intersection rules.

This paper has been organized as follows. In section 2, we present the fields, couplings, field equations and Bianchi identities of (1,0) superconformal theories with hyper-multiplets. In section 3,
we give the solutions to the hyperini KSEs. In section 4, we construct new string and 3-brane solutions and in section 5, we give our conclusions.

\newsection{(1,0) superconformal theory and KSEs}

\subsection{Fields and KSEs}

The (1,0) superconformal models constructed in \cite{ssw, ssw3} have vector, tensor and hyper-multiplets as well as   appropriate higher form fields which appear  in Stuckelberg-type of couplings.
The field content of the vector multiplets is $(A_\mu^r, \lambda^{ir}, Y^{ijr})$, where $r$ labels the different vector multiplets and $i, j = 1,2$ are the $Sp(1)$ R-symmetry indices, $A_\mu^r$ are 1-form gauge potentials, $\lambda^{ir}$ are symplectic Majorana-Weyl spinors and $Y^{ijr}$ are auxiliary fields.
The field content of the tensor multiplets is  $(\phi^I, \chi^{iI}, B_{\mu\nu}^I)$, where $I$ labels the different tensor multiplets, $\phi^I$ are scalars, $\chi^{iI}$ are symplectic Majorana-Weyl spinors, of opposite chirality from those
of the vector multiplets, and $B_{\mu\nu}^I$ are the 2-form gauge potentials.
The field content of the hyper-multiplets are $(q^\alpha , \psi^a)$, where $q^\alpha$ are the ``hyper-scalars'', which are maps from the spacetime to a  hyper-K\"ahler cone\footnote{Supersymmetry
requires that the hyper-scalars take values on a hyper-K\"ahler manifold. In addition superconformal symmetry requires that the hyper-K\"ahler manifold
admits a homothetic motion associated with a potential making the hyper-K\"ahler manifold locally a hyper-K\"ahler cone.}, and $\psi^a$ are symplectic Majorana-Weyl spinors of the same chirality as $\chi^{iI}$.

The field strengths of the 1- and 2-form gauge potentials associated with the vector and tensor multiplets are
\be
\mathcal{F}_{\mu\nu}^r &\equiv& 2\partial_{[\mu}A_{\nu]}^r - f_{st}{}^rA_\mu^s A_\nu^t + h_I^rB_{\mu\nu}^I~,\\
\mathcal{H}_{\mu\nu\rho}^I &\equiv& 3D_{[\mu}B_{\nu\rho]}^I + 6d_{rs}^I A_{[\mu}^r\partial_\nu A_{\rho]}^s - 2f_{pq}{}^s d_{rs}^I A_{[\mu}^r A_\nu^p A_{\rho]}^q + g^{Ir}C_{\mu\nu\rho r}~,
\ee
respectively, where $f_{rs}{}^t$   $h_I^r, g^{Ir}$ and $d_{rs}^I=d_{(rs)}^I$ are coupling constants, and $C_{\mu\nu\rho r}$ are  three-form gauge potentials introduced
via a St\"uckelberg-type of coupling. In addition,
\be
D_\mu \Lambda^s \equiv \partial_\mu \Lambda^s+ A_\mu^r(X_r)_t{}^s \Lambda^t~,~~~~D_\mu \Lambda^I \equiv \partial_\mu \Lambda^I+ A_\mu^r(X_r)_J{}^I \Lambda^J~,
\ee
where $X_r$ are given by
\bea
 (X_r)_t{}^s=-f_{rt}{}^s+d^I_{rt} h^s_I~,~~~(X_r)_J{}^I=2 h^s_J d^I_{rs}-g^{Is} b_{Jsr}~.
 \eea
The various coupling satisfy a long list of restrictions required by gauge invariance and for these models to have an action which is given in \cite{ssw}, see also \cite{ap3} for a summary,  and it will not be repeated here.
In particular, these models are described by an action provided there is a maximally split signature metric\footnote{Since the metric is maximally split,
the kinetic energy of some of the fields is negative which may lead to ghosts in the spectrum. This is an issue affecting this class of theories.}  $\eta_{IJ}$ such that
\bea
g^{Ir}=\eta^{IJ} h^r_I~,~~~d^I_{rt}={1\over2}\eta^{IJ}b_{Jrt}~.
\eea
From now on, the indices $I,J$ are raised and lowered with $\eta$.

To couple hyper-multiplets to the above system \cite{ssw3}, one assumes that the hyper-K\"ahler cone, which is the target space of hyper-multiplet scalars,  admits tri-holomorphic isometries generated by the
vector fields $X_{(\mathfrak{m})}=X^\alpha_{(\mathfrak{m})}\partial_\alpha$. Typically only some of the vector multiplets will be gauged. For this introduce the embedding tensor $\theta^\mathfrak{m}_r$
and define
\be
A^{\mathfrak{m}} = A^r \theta_r{}^{\mathfrak{m}}~,~~~ \lambda^{\mathfrak{m}} = \lambda^r\theta_r{}^{\mathfrak{m}}~,~~~Y^{\mathfrak{m}}_{ij} = Y^r_{ij} \theta_r{}^{\mathfrak{m}}~,
\label{theta1}
\ee
where for consistency with the gauge transformations
\be
h^r{}_I\theta_r{}^{\mathfrak{m}} = 0~,~~~~~~f_{rs}{}^t\theta_t{}^{\mathfrak{m}} = \theta_r{}^{\mathfrak{n}}\theta_s{}^{\mathfrak{p}}f_{\mathfrak{n}\mathfrak{p}}{}^{\mathfrak{m}}~,
\label{theta2}
\ee
and where $[X_{(\mathfrak{n})}, X_{(\mathfrak{p})}]=-f_{\mathfrak{n}\mathfrak{p}}{}^{\mathfrak{m}} X_{(\mathfrak{m})}$.
The KSEs of the model, which are the vanishing conditions for the supersymmetry transformations of the fermions evaluated at the locus where all fermions vanish, are
\bea
\delta \lambda^{ir} &=& \frac{1}{8}\mathcal{F}_{\mu\nu}^r\Gamma^{\mu\nu} \epsilon^i - \frac{1}{2}Y^{ijr}\epsilon_j + \frac{1}{4}h_I^r\phi^I\epsilon^i=0~,\cr
\delta \chi^{iI} &=& \frac{1}{48}\mathcal{H}_{\mu\nu\rho}^I \Gamma^{\mu\nu\rho}\epsilon^i + \frac{1}{4}D_\mu\phi^I \Gamma^\mu \epsilon^i =0~,\cr
\delta \psi^a &=& \frac{1}{2}D_\mu q^\alpha \Gamma^\mu \epsilon_i E^{ia}{}_{\alpha} =0~,
\label{kse}
\eea
where
\be
D_\mu q^\alpha = \partial_\mu q^\alpha -A^{\mathfrak{m}}_\mu X^\alpha_{(\mathfrak{m})}~.
\ee
In addition, $E^\alpha_{ia}$ is the symplectic frame of the hyper-K\"ahler cone, ie the hyper-K\"ahler metric and hypercomplex structure   are given as
\bea
g_{\alpha\beta}= \epsilon_{ij} \epsilon_{ab} E^{ia}{}_{\alpha} E^{jb}{}_{\beta}~,~~~(I_\tau)^\alpha{}_\beta=-i\, (\sigma_\tau)^i{}_j\, \delta^a{}_b\, E_{ia}^\alpha\, E^{jb}_\beta~,~~~
\eea
where $\epsilon_{ij}$ and $\epsilon_{ab}$ are the symplectic (fundamental) forms of $Sp(1)$ and $Sp(n)$, respectively,  and $\sigma_\tau,  \tau=1,2,3$ are the Pauli matrices. In analogy with similar variations in 6-dimensional (1,0) supergravity, we refer to these KSEs as the gaugini, tensorini and hyperini KSEs, respectively.

The Lagrangian for these theories consist of two parts. One  part, $\mathcal{L}_{VT}$,  involves the vector and tensor multiplets, and the second part, $\mathcal{L}_H$, contains the hyper-multiplets. These two parts are independently supersymmetric and the supersymmetry transformation of the vector multiplets used in the coupling of the hyper-multiplets in $\mathcal{L}_H$ is obtained by contraction with the embedding tensor.

\subsection{Field equations }

The field equations of the system are
\bea
D^\mu D_\mu \phi^I &=& -\frac{1}{2}d_{rs}^I(\mathcal{F}_{\mu\nu}^r\mathcal{F}^{\mu\nu s} - 4Y_{ij}^rY^{ijs}) - 3d_{rs}^Ih_J^rh_K^s\phi^J\phi^K~,\label{eqa}
\cr
b_{Irs}Y_{ij}^s\phi^I &=&{1\over2\lambda} \theta_r{}^\km\, \mu_{\km ij} ~,\label{eqb}
\cr
b_{Irs}\mathcal{F}_{\mu\nu}^s\phi^I &=&\frac{1}{4!}\epsilon_{\mu\nu\lambda\rho\sigma\tau}\mathcal{H}_r^{(4)\lambda\rho\sigma\tau}~,
\cr
g_{\a\b} \nabla_\mu D^\mu q^\b&=&-Y^\km_{ij} \partial_\a \mu^{ij}_{(\km)}~,
\label{eqc}
\eea
where
\bea
\nabla_\mu D^\mu q^\a=\partial_\mu D^\mu q^\a+\Gamma^\a_{\b\g} D^\mu q^\b  D_\mu q^\g-\partial_\b X_{(\km)}^\a  \theta^\km_r A_\mu^r D^\mu q^\b~,
\eea
$\lambda$ is a constant, and $\mu_{(\mathfrak{m})\tau}$,
\bea
X_{(\mathfrak{m})}^\beta (\omega_\tau)_{\beta\alpha}=-\partial_\alpha\mu_{(\mathfrak{m})\tau}~,~~~(\omega_\tau)_{\alpha\beta}=g_{\alpha\gamma}(I_\tau)^\gamma{}_\beta~,
\eea
are the moment maps.
Observe that generically the theory has a cubic scalar field interaction and so the potential term is not bounded
from below. These field equations are also supplemented with
the Bianchi identities
\bea
D_{[\mu}\mathcal{F}_{\nu\rho]}^r &=& \frac{1}{3}h_I^r\mathcal{H}_{\mu\nu\rho}^I~,
\cr
D_{[\mu}\mathcal{H}_{\nu\rho\sigma]}^I &=& \frac{3}{2}d_{rs}^I\mathcal{F}_{[\mu\nu}^r\mathcal{F}_{\rho\sigma]}^s + \frac{1}{4}g^{Ir}\mathcal{H}_{\mu\nu\rho\sigma r}^{(4)}~,
\cr
D_{[\mu}\mathcal{H}^{(4)}_{\nu\lambda\rho\sigma]r}&=&-4d_{Irs} {\mathcal F}^s_{[\mu\nu} {\mathcal H}^I_{\lambda\rho\sigma]}+{1\over5}\theta_r{}^\km {\cal H}^{(5)}_{\km \mu\nu\lambda\rho\sigma}~,
 \label{idh}
\eea
where $\mathcal{H}_{\mu\nu\rho\sigma r}^{(4)}$ is the field strength of the 3-form, and the duality relations
\bea
{1\over 5!} \epsilon_{\mu\nu\rho\lambda\sigma \tau} \theta_r{}^\km \mathcal {H}^{(5)\nu\rho\lambda\sigma \tau}_\km =(X_r)_{IJ} \phi^I D_\mu \phi^J+{2\over\lambda} \theta_r{}^\km X_{(\km)}{}_\alpha D_\mu q^\alpha~,
\eea
ie the 5-form field strength is dual to the hyper-scalars.

\section{Solution Hyperini KSEs}

\subsection{Spinorial geometry}

The inclusion of the hyper-multiplets does not alter the  first two KSEs in (\ref{kse}).  Because of this, the solution of these equations  described  in \cite{ap3}, using the spinorial geometry technique of \cite{uggp}, still applies and the results will not be repeated here.

It remains to solve the hyperini KSEs in (\ref{kse}). This has a  similar structure to that of  the hyperini  KSEs which has been solved  in the context of
6-dimensional (1,0) supergravity \cite{ap1}. Because of this we shall not give details of the derivation. Instead, we shall state the result of the spinorial geometry calculation and then
 re-express it in a way which allows for an improved interpretation of the conditions that arise.  This is instrumental in the description of 3-brane solitons in some of the
 (1,0) superconformal models in section 4.

The Killing spinors and their isotropy groups in $Spin(5,1)\cdot Sp(1)$ that arise in the solution  of all KSEs in (\ref{kse}) are summarized in table 1. The isotropy groups are either compact or non-compact and we shall use this to distinguish solutions with the same supersymmetry. Note that
solutions with $N=3$ supersymmetry arise only when hyper-multiplets are present.

\begin{table}[ht]
 \begin{center}
\begin{tabular}{|c|c|c|}
\hline
$N$&${\mathrm{Isotropy ~Groups}}$  & ${\mathrm{Killing ~Spinors}}$ \\
\hline
\hline
$1$  & $Sp(1)\cdot Sp(1)\ltimes \bH$ & $1+e_{1234}$\\
\hline
$2$  & $(U(1)\cdot
Sp(1))\ltimes\bH$ & $1+e_{1234}~, ~i(1-e_{1234})$\\
\hline
$3$  & $
Sp(1)\ltimes \bH$ & $1+e_{1234}~, ~i(1-e_{1234})~,~e_{12}-e_{34}~$ \\
\hline
$4$  & $
Sp(1)\ltimes \bH$ & $1+e_{1234}~, ~i(1-e_{1234})~,~e_{12}-e_{34}~,~i(e_{12}+e_{34})$\\
\hline
\hline
$2$  & $
Sp(1)$ & $1+e_{1234}~, ~e_{15}+e_{2345}$\\
\hline
$4$  & $
U(1)$ & $1+e_{1234}~,~i(1-e_{1234})~, ~e_{15}+e_{2345}~,~ i(e_{15}-e_{2345})$\\
\hline
\end{tabular}
\end{center}
\caption{\small
The first column gives the number of invariant spinors, the second column the associated isotropy groups and the third column representatives of the invariant spinors.
The isotropy group of more than 4 linearly independent spinors is the identity.}
\end{table}

A detailed analysis of how the Killing spinors in table 1 are selected is given in \cite{ap1}.  Here we shall use this selection and substitute it in the hyperini KSEs to find the conditions
that arise on the fields. For this note that in the context of spinorial geometry the lowering of the $Sp(1)$ indices on the supersymmetry parameter $\epsilon$ that appears in the hyperini KSEs
is implemented by
\be
\epsilon_1 = -\epsilon^2~, ~~~\epsilon_2 = \Gamma_{34}\epsilon^1~,
\label{hypspin}
\ee
where $(\epsilon^i)=(\epsilon^1, \epsilon^2)$ and $\epsilon^1, \epsilon^2$ are the two Weyl spinors that we used to construct the symplectic-Majorana representation, and the gamma matrices
$\Gamma_3, \Gamma_4$ are along two auxiliary directions which arise because we have identified the 6-dimensional symplectic-Majornana spinors with the $SU(2)$ invariant spinors of the positive chirality  Majorana-Weyl
representation of $Spin(9,1)$. For the spinor notation we use and other details see \cite{ap1, ap3}.

\subsection{N=1}
The Killing spinor can be chosen as $\epsilon = 1 +e_{1234}$ which is invariant under the subgroup $Sp(1)\cdot Sp(1)\ltimes \bH$ of $Spin(5,1)\cdot Sp(1)$. In this case, $(\epsilon^i)=(\epsilon^1, \epsilon^2)=
(1, e_{1234})$, and using the relations in (\ref{hypspin}) we have $(\epsilon_i)=(\epsilon_1, \epsilon_2)=(-e_{1234}, e_{34})$. Substituting  these into the hyperini KSEs, we find
\be
D_+q^\alpha E^{ia}{}_\alpha = 0~,~~~ -D_1 q^\alpha E^{1a}{}_\alpha + D_{\bar{2}}q^\alpha E^{2a}{}_\alpha = 0~,~~~D_{2}q^\alpha E^{1a}{}_\alpha + D_{\bar{1}}q^\alpha E^{2a}{}_\alpha = 0~,
\la{n1c}
\ee
where in the evaluation of the above conditions the 6-dimensional spacetime decomposes into two light-cone directions and four transverse directions which are written in terms of complex coordinates, eg the Minkowski  spacetime metric is written as
\bea
ds^2= \eta_{\mu\nu} dx^\mu dx^\nu=2 dx^+ dx^-+ \delta_{mn} dx^m dx^n=2 (dx^+ dx^-+  dz^1 dz^{\bar 1}+dz^2 dz^{\bar 2})~.
\eea
The first condition simplifies to
\be
D_+q^\alpha = 0~,
\ee
ie the hyper-scalars are covariantly constant along one of the lightcone directions and so in the gauge $A_+^\mathfrak{m}=0$  do not depend on $x^+$.  The remaining two conditions in (\ref{n1c})  can be written in a more covariant form as
\be
(\tau^m)^i{}_j D_m q^\alpha E^{ja}{}_\alpha = 0~,
\ee
where $(\tau^m)=( -i\sigma_\tau, 1_{2\times 2})$, $m=1,2,3, 4$ and $\sigma_\tau$ are the Pauli matrices, or equivalently in a coordinate basis as
\bea
(\mathfrak{I}^m)^\alpha{}_\beta D_{m} q^\beta=0~,
\eea
where $(\mathfrak{I}^m)= (I_\tau, 1_{4n\times 4n})$.

\subsection{N=2 Non-Compact}
From table 1, the Killing spinors can be chosen as $ 1+e_{1234}$ and $i(1-e_{1234})$. Using these in the hyperini KSEs, we find the following conditions
\be
D_+q^\alpha = 0~, ~~~ D_1q^\alpha E^{1a}{}_\alpha = D_2q^\alpha E^{1a}{}_\alpha = 0~,~~~ D_{\bar{1}}q^\alpha E^{2a}{}_\alpha = D_{\bar{2}}q^\alpha E^{2a}{}_\alpha = 0~.
\ee
The last two conditions in the above equation  can be rewritten as a Cauchy-Riemann (CR) type of equation
\be
D_m q^\alpha(I_3)^{ia}{}_{jb}E^{jb}{}_\alpha = J^{n}{}_{m} D_{n}q^\alpha E^{ia}{}_{\alpha}~,
\ee
where $(I_3)^{ia}{}_{jb} = (-i\sigma_3)^i{}_j\delta^a{}_b$ and $J^n{}_m = (i\delta^{\mathrm{s}}{}_{\mathrm {r}}, -i\delta^{\bar{\mathrm{s}}}{}_{\bar{\mathrm {r}}})$, $\mathrm {r},\mathrm{s}=1,2$.  Therefore in the absence of gauge
fields, the above condition becomes the CR equation and $q$ is a holomorphic map from the transverse space to the light-cone to the hyper-K\"ahler cone with respect to the pair of complex structures\footnote{The choice of complex structure
on the hyper-K\"ahler cone depend on the choice of representatives for the Killing spinors. For a generic choice, the complex structure $I_3$ should be replaced
with a linear combination of all three complex structures.} $(J, I_3)$.

\subsection{N=2 Compact}
The two Killing spinors can be chosen as $ 1+e_{1234}$ and $e_{15} + e_{2345}$. The conditions imposed by the hyperini KSEs  evaluated on $e_{15} + e_{2345}$ can be written
\be
D_-q^\alpha E^{ia}{}_\alpha = 0~, ~~~ -D_2q^\alpha E^{1a}{}_\alpha + D_1q^\alpha E^{2a}{}_\alpha = 0~, ~~~ D_{\bar{1}}q^\alpha E^{1a}{}_\alpha + D_{\bar{2}}q^\alpha E^{2a}{}_\alpha = 0~.
\ee
Combining these  conditions with those associated with $ 1+e_{1234}$, we find that the hyper-scalars satisfy
\be
D_{\underline r}q^\alpha =0~,~~~~~~{\underline r}=-, +, 1~,
\ee
and
\be
D_{r'}q^\alpha = -\epsilon_{r'}{}^{s't'}(K_{s'})^\alpha{}_\beta D_{t'}q^\beta~,~~~r', s', t'=2,3,4~,
\label{cn2}
\ee
where we have made a 3+3 (real) split of the spacetime  and $(K_{s'})^{ia}{}_{jb} = -i(\sigma_{s'-1})^i{}_j\delta^a{}_b$ with $\epsilon_{234}=1$.  Therefore the hyper-scalars are covariantly constant along the first three directions of the spacetime and obeys (\ref{cn2}) along the other three.

\subsection{N=3 Non-Compact}
The three Killing spinors  are $1+e_{1234}$, $i(1-e_{1234})$ and $e_{12}-e_{34}$. The conditions imposed by the hyperini KSEs on the hyper-scalars  are
\be
D_+q^\alpha  = 0~, ~~~ D_1q^\alpha E^{1a}{}_\alpha = D_2q^\alpha E^{1a}{}_\alpha = 0~,~~~ D_{\bar{1}}q^\alpha E^{2a}{}_\alpha = D_{\bar{2}}q^\alpha E^{2a}{}_\alpha = 0~, \nonumber \\
D_{\bar{1}} q^\alpha E^{1a}{}_\alpha - D_2q^\alpha E^{2a}{}_\alpha = 0~,~~~D_{\bar{2}}q^\alpha E^{1a}{}_\alpha + D_{1}q^\alpha E^{2a}{}_\alpha = 0~.
\ee
As in the previous non-compact cases, the hyper-scalars are covariantly constant along one of the lightcone directions. The remaining conditions can be written as
\bea
(J_\tau)^m{}_n D_m q^\alpha=(I_\tau)^\alpha{}_\beta D_n q^\beta~,~~~\tau=1,2,3~,
\la{quatr}
\eea
for an appropriate choice of a hypercomplex structure $J_\tau$ in the directions transverse to the light cone with $J_3=J$. Therefore in the absence of gauge couplings, the hyper-scalars are quaternionic maps\footnote{Clearly, the directions transverse to the lightcone can be identified with the quaternions $\bH$. If the Obata curvature vanishes, then it is possible to introduce quaternionic coordinates on the hyper-K\"ahler cone. In such a case $q$'s can be written as  quaternions ${\bf q}$ and (\ref{quatr}) implies that ${\bf q}={\bf q}({\bf x}, x^-)$, ${\bf x}\in \bH$.}  from the
directions transverse to the lightcone to the hyper-K\"ahler cone.
\subsection{N=4 Non-Compact}
The four Killing spinors can be chosen as $1+e_{1234}$, $i(1-e_{1234})$, $e_{12}-e_{34}$ and $i(e_{12}+e_{34})$. The only non-vanishing component of the hyper-scalars is
$D_-q^\alpha$,
i.e. in the absence of gauge fields the hyper-scalars depend only on the light-cone direction $x^-$.

\subsection{N=4 Compact}
The four Killing spinors can be chosen as $1+e_{1234}$, $i(1-e_{1234})$,  $e_{15}+e_{2345}$ and $i(e_{15}-e_{2345})$. The conditions imposed on the hyper-scalars  from the hyperini KSEs are
\bea
D_\ur q^\alpha = 0~, ~~~~~~ \ur = -,+,1, \bar{1}~,\cr
D_2q^\alpha E^{1a}{}_\alpha = D_{\bar{2}}q^\alpha E^{2a}{}_\alpha = 0~.
\la{n4comp}
\eea
Therefore there is a 4+2 (real) split of the spacetime and  the hyper-scalars are covariantly constant along the first four directions. In the remaining two directions, the hyper-scalars satisfy a CR type of equation
\bea
J^{r'}{}_{s'} D_{r'} q^\alpha =  (I_3)^{\alpha}{}_{\beta}  D_{s'} q^\beta~,
\la{n4compb}
\eea
where $J^{r'}{}_{s'}=(i \delta^2{}_2, -i \delta^{\bar 2}{}_{\bar 2})$.

\subsection{Maximal Supersymmetry}

All solutions of the hyperini KSEs with more than 4 Killing spinors are maximally supersymmetric, ie they preserve all 8 supersymmetries.
In addition, the hyperini KSEs imply that for maximally supersymmetric backgrounds the hyper-scalars are covariantly constants, ie
\bea
D_\mu q^\alpha=0~.
\eea
This concludes the description of solutions of the hyperini KSEs.

\newsection{Brane solitons}

\subsection{Self-dual string solitons}

\subsubsection{A class of models}

A large class of models has been constructed in \cite{ssw, ssw3}  by considering a Lie algebra $\mathfrak{g}$ and a representation ${\cal R}$. The bosonic fields of the
 vector and tensor multiplets are chosen as
\bea
A^r=(A^\km, A^A)~,~~~Y^r=(Y^\km, Y^A)~,~~~B^I=(B^A, B_A)~,~~~\phi^I=(\phi^A, \phi_A)~,
\eea
ie $A$ and $Y$ take values in $\mathfrak{g}\oplus {\cal R}$ while $B$ and $\phi$ take values in  ${\cal R}\oplus {\cal R}^* $.  Moreover the
non-vanishing couplings are chosen as
\bea
&&\eta^A{}_B=\eta_B{}^A=\delta^A_B~,~~~h^B{}_A=g_A{}^B=\delta^B_A~,~~~f_{\km A}{}^B=-{1\over2} (T_\km)_A{}^B~,~~~f_{\km\kn}{}^\kp~,
\cr
&&d^B{}_{\km A}={1\over2} b^B{}_{A\km}={1\over2} b^B{}_{\km A}={1\over2} (T_\km)_A{}^B~,~~~d_{ABC}=d_{(ABC)}=b_{BCA}~,~~~
\cr
&&d_{AB\km}=d_{(AB)\km}={1\over2} b_{AB\km}={1\over2} b_{A\km B}~,~~~d_{A\km\kn}~,~~~
b_{A(\km\kn)}=2 d_{A(\km\kn)}~,~~~\theta_\km{}^\kn=\delta_\km{}^\kn~,
\eea
where $T_\km$ are the representation matrices of $\mathfrak{g}$ in ${\cal R}$. These solve all the constraints on the couplings imposed on these models
provided that $d_{\km AB}, d_{\km\kn A}$ and $d_{ABC}$ are invariant under the action of $\mathfrak{g}$.

\subsubsection{Self-dual string solitons from instantons}

Motivated from the M-brane intersection rules, we shall seek   self-dual string solitons in the  class of models described in the previous section which preserve $1/2$ of the supersymmetry. The relevant class of
supersymmetric backgrounds for self-dual string solitons are those with 4 Killing spinors that have isotropy group $Sp(1)\ltimes \bH$ in table 1.  The conditions
on the field of the vector and tensor multiplets are given in \cite{ap3} and in section (3.6) for the hyper-scalars. Similar solutions have been found in \cite{ap3} for another class of
models. The string soliton on a single M5-brane has been found in \cite{howe} and it is singular at the position of the brane.

To solve the supersymmetry conditions, Bianchi identities and field equations, suppose that the fields have support on 4-directions transverse to the lightcone which are identified with the world-sheet of the string. In addition  choose
\bea
\mathcal{F}^r=(\mathcal{F}^\km,0)~,~~\mathcal{H}^I=(0, \mathcal{H}_A)~,~~\phi^I=(0, \phi_A)~,~~\mathcal{H}_r^{(4)}=Y^r=\mathcal{H}^{(5)}=0~,
\eea
with $\mathcal{F}^r$ purely magnetic.
We focus on models for which the only non-vanishing coupling constants with all indices lowered are $b_{A\km\kn}, d_{A\km\kn}$.  In addition we assume that either the model is not coupled
to hyper-multiplets or if it couples the hyper-scalars are at a maximally supersymmetric vacuum for consistency, ie the gauging and the hyper-K\"ahler cone have chosen such that there is a value $q=q_0$ such that
\bea
\mu_\km(q_0)=0~,~~~\partial_\alpha\mu_\km(q_0)=0~.
\eea
For the flat hyperk\"ahler cone, such a value is $q_0=0$ or any other fixed point of rotational isometries that are gauged. In either case, the contribution
from the hyper-multiplets decouples.

The remaining  non-trivial Bianchi identities and field equations that one has to demonstrate are
\bea
D_{[m}\mathcal{F}^\km_{n\ell]}=0~,~~~b^B{}_{A\kn }\mathcal{F}^\kn_{m\ell} \phi_B=0~,~~~
d^B{}_{A\mathfrak{n}} \mathcal{F}^\kn_{[\mu\nu} \mathcal{H}_{\lambda\rho\sigma ]B}=0~,~~~\la{cona}
\eea
and
\bea
D_{[\mu} \mathcal {H}_{\nu\rho\sigma] A}={3\over2} d_{A\km\kn} \mathcal{F}^\km_{[\mu\nu} \mathcal{F}^\kn_{\rho\sigma]}~,  D_m D^m\phi_A=-{1\over2} d_{A\km\kn}  \mathcal{F}^\km_{mn}  \mathcal{F}^{\kn mn}~.
\la{conb}
\eea

These conditions can be solved provided that ${\cal R}$ can be decomposed as ${\cal R}=I\oplus {\cal R}'$, where $I$ is a trivial representation of $\mathfrak{g}$ and take that
$\phi_A$ and $\mathcal{H}_A$ vanish unless they lie along the trivial representation, and denote the non-vanishing fields with $\phi_0$ and $\mathcal{H}_0$, respectively. Such a choice will solve the last two conditions in (\ref{cona}) as $T_\km$ vanishes
along the trivial representation. The first condition in (\ref{cona}) is solved by identifying $\mathcal{F}^\km$ with the field strength of a gauge field with Lie algebra $\mathfrak{g}$.

It remains to solve the conditions in (\ref{conb}). First observe that $D_m D^m\phi_0=\partial_m\partial^m \phi_0$, and similarly on $\mathcal{H}_0$, and identify $d_{0\km\kn}$ with a bi-invariant metric on $\mathfrak{g}$.   Next set
\bea
\mathcal{H}_0=- \partial_m\phi_0 dx^-\wedge dx^+\wedge dx^m+{1\over3!}
\partial_\ell\phi_0\,\epsilon^\ell{}_{m_1m_2m_3}\,\,dx^{m_1}\wedge dx^{m_2}\wedge dx^{m_3} ~,~~
\la{exh}
\eea
Then recall that the KSEs for $Sp(1)\ltimes \bH$ invariant spinors
require that $\mathcal{F}$ is an anti-self dual instanton. Because of this and (\ref{exh}), the second condition in (\ref{conb}) implies the first.  Finally, the last condition
in (\ref{conb}) is solved because the Pontryagin form of instantons can be written as the Laplacian on a scalar \cite{osborn}.  A similar calculation has been
done for the self-dual string solution in \cite{ap3} and so no further details will be presented.  Notice though that this new class of self-dual string solutions
is more general than that of \cite{ap3} as there is no restriction on the gauge group. In addition for generic values of instanton moduli space, all the string solutions
are smooth and  the string charge is related to the instanton number, see also \cite{halfhet} for more details.

\subsection{3-branes}

Motivated from the M-brane intersection rules which state that two M5-branes intersect on a 3-brane, we shall describe a class of models which exhibit 3-brane solitons. These are those for which all the potentials vanish and the only active fields are those of the hyper-multiplets.
Moreover, the hyper-multiplet scalars depend only on the two transverse directions of the 3-brane soliton. First to identify the models with 3-brane solitons
suppose that the hyper-multiplets are not gauged, ie the embedding tensor $\theta=0$.  Moreover set all the fields apart from the hypermultiplet scalars $q$
and $\mathcal{H}^{(5)}$ equal to zero. The only non-trivial conditions that have to be satisfied to construct solutions are the field equations for $q$
and the hyperini KSEs.

To solve the hypernini KSEs, we shall take the case with 4 supersymmetries and compact isotropy group.  The relevant equations are given in (\ref{n4comp}) and (\ref{n4compb}).  From the KSEs, the hyper-multiplet scalars do not depend on four directions as expected for a 3-brane soliton and (\ref{n4compb})  is a   Cauchy-Riemann
equations implying that $q$ is a holomorphic curve into the hyper-K\"ahler cone.  In addition, the field equation for the $q$'s  is automatically satisfied.

A similar argument based on the supersymmetry conditions in section 3.3 leads to the existence of a string soliton preserving 1/4 of supersymmetry supported by a homolorphic surface embedded 
into the hyper-K\"ahler cone. Such solitons are expected to exist as they can be  associated with a triple M5-brane intersection on a string. Similarly, the ``quaternionic'' solitons
of section 3.5 can be associated with M-brane intersections at angles.

\newsection{Concluding Remarks}

Combining the results of \cite{ap3} with those of this paper, we have solved the KSEs of the (1,0) superconformal models of \cite{ssw, ssw3} in all cases and identified  the fractions of supersymmetry preserved.  The theories have a large number of solitons. Here we have presented new self-dual string solutions and  3-brane solutions which are expected to exist
because of the M-brane intersection rules. The former  supported by instantons with arbitrary gauge group, they are smooth at a generic point of the instanton  moduli space and the string charge is related to the instanton number. The latter
are  holomorphic curves from the 6-dimensional spacetime  into the hyper-K\"ahler cone of the hyper-multiplets.

Although it may appear that the techniques we have used to solve the KSEs apply to a particular class of models, this is not the case. In fact, the same method can be applied
to solve the KSEs of any (1,0) supersymmetric theory in 6 dimensions as it has been demonstrated in \cite{ap1}.

It is clear that there are many more supersymmetric solutions like the string solitons mentioned in the previous section associated with triple M5-brane intersections that preserve $1/4$ of supersymmetry. This can be seen either by directly inspecting the solutions of the KSEs  we have presented or by continuing to explore the analogy with the M-brane intersection rules. The KSEs that we have solved are the most general ones for models with (1,0) supersymmetry in six dimensions. So, there is now a systematic way to find all solitons of these superconformal theories and to explore
their applications.

\vskip 1.0cm
\noindent{\bf Acknowledgements} \vskip 0.1cm
\noindent
MA has been partially supported by a STFC studentship.
GP is partially supported by the STFC rolling grant ST/G000/395/1.
\vskip 0.5cm
\setcounter{section}{0}
\setcounter{subsection}{0}

\end{document}